\documentclass[aps,prb,twocolumn,groupedaddress]{revtex4-2}

\usepackage{graphicx}
\usepackage{physics}

 \newcommand{\change}[1]{#1}

\begin{document}

\preprint{Submitted to PRB}

\title{Transport-Induced Decoherence of the Entangled Triplet Exciton Pair}

\author{Gerald Curran III, Luke J. Weaver, Zachary Rex, Ivan Biaggio}
\affiliation{Department of Physics, Lehigh University, Bethlehem PA 18018, USA}
 
\date{\today}

\begin{abstract}
Decoherence effects for entangled triplet exciton pairs in organic molecular crystals are analyzed for the case when excitons can hop between inequivalent lattice sites. The fluorescence quantum beats caused by quantum interference upon triplet-triplet recombination into an emissive singlet state are predicted as a function of hopping time and magnetic field based on a Monte Carlo analysis. Depending on exciton hopping rates, it is possible to have   global decoherence and complete suppression of fluorescence quantum beats in the limit of zero magnetic field, and to have quantum beats that decay at different rates depending on magnetic field strength.
\end{abstract}


\maketitle

\section{Introduction}
In crystals where excitons can localize at inequivalent lattice sites, the Hamiltonian and the stationary states that describe exciton pairs vary as excitons diffuse.

In the case of pairs of triplet excitons created by singlet exciton fission, their spin-entanglement and the time evolution of their spin wavefunction affect their probability of recombining into a singlet state, which can result in quantum interference and quantum beats in short-pulse induced fluorescence. \cite{Chabr81,Funfschilling85b,Wolf18,Curran24}

However, for quantum interference in the exciton pair to lead to observable quantum beats it is necessary to maintain global coherence in the triplet-pair population. While quantum beats can be easily observed in tetracene crystals \cite{Funfschilling85b,Chabr81,Burdett12}, in rubrene crystals they can only be detected in the presence of a carefully aligned magnetic field \cite{Wolf18,Curran24}. The reason is a global decoherence effect associated with triplet transport, which was analyzed in the high-field limit in Ref.~\onlinecite{Curran24}. 

In this work we focus on the effect of excitons hopping between inequivalent sites at zero magnetic field and moderate magnetic fields, highlighting how exciton transport can lead to global decoherence in a population of triplet exciton pairs created by singlet exciton fission, the suppression of the quantum beat signal at zero field, and the decay of quantum beats amplitude at moderate magnetic fields, even in the absence of the transport-induced dephasing effect discussed in Ref.~\onlinecite{Curran24}. Examples of decoherence due to hopping between inequivalent sites were discussed for individual triplet excitons and other systems in Refs.~\onlinecite{Berk83,Baker12,Weiss17,Palmer25}. In this work we specifically develop a complete model and computational method to predict transport-induced decoherence for a population of entangled triplet exciton pairs in any magnetic field. 

\change{One can generally describe decoherence effects in terms of \emph{global decoherence}, which affects a population of states, and of \emph{local decoherence}, which affects the wavefunction of individual states. In our analysis of the time-evolution of the spin wavefunction of triplet-exciton pairs, we focus on global decoherence as the dominant cause of the suppression of observable quantum beats (as seen in the example of rubrene). Local decoherence, as for example caused by randomization of the spin orientation of individual triplet excitons because of spin-orbit coupling \cite{Yu12}, generally occurs on a longer time-scale and does not influence the effects discussed here.}

\begin{figure}
\includegraphics[width=0.9\columnwidth]{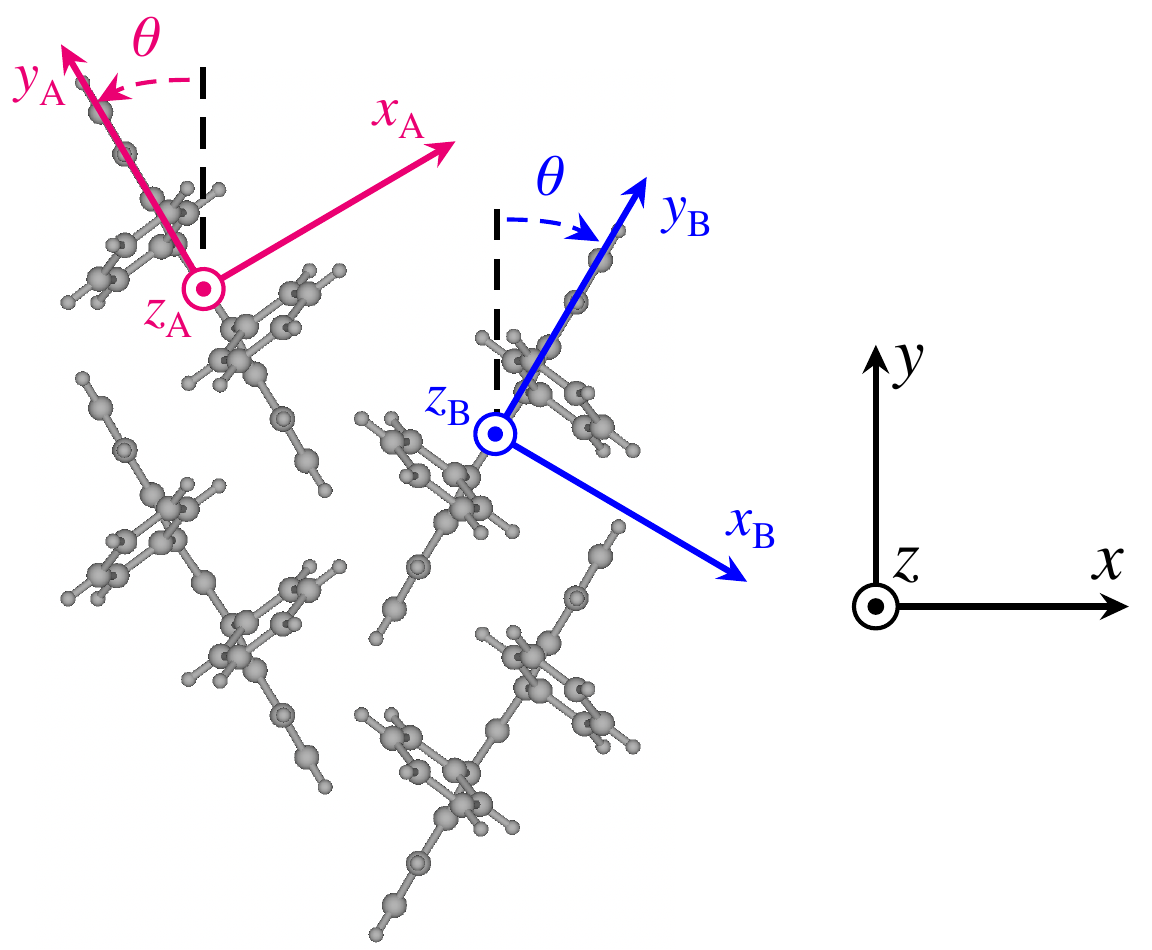}
\caption{Schematic of inequivalent site coordinate systems and global crystal coordinate system in the example of orthorhombic rubrene crystals. The $y_A$ and $y_B$ axes of each site are along the tetracene backbone of the rubrene molecules, and each site's coordinate system is rotated around a shared $z$-axis by the herringbone angle $\pm \theta$. In orthorhombic rubrene crystals, $\theta = 31^\circ$ \cite{Jurchescu06}, with crystal axes labels $a=x,b=y,c=z$ in the convention of Ref.~\onlinecite{Wolf18,Curran24} and $a=z,b=y,c=x$ in the convention of Ref.~\onlinecite{Jurchescu06} } \label{Axes}
\end{figure}

We start by reviewing the spin wavefunction of the entangled triplet-pair and how the corresponding Hamiltonian is modified when the two excitons in a pair reside on different inequivalent sites. We then develop a model that accounts for hopping rate and magnetic field strength and that predicts the emergence and persistence of fluorescence quantum beats. The analysis is general and can be applied to any crystal structure where excitons can localize at two inequivalent sites in a unit cell, but we will provide examples based on rubrene orthorhombic single crystals. All results that we will give below for the example of rubrene are calculated for a $\theta = 31^\circ$ herringbone angle (See Fig.~\ref{Axes}) and molecular zero-field splitting parameters $D = 0.0555$~cm$^{-1}$ and $E = - 0.0040$~cm$^{-1}$ as determined for the rubrene molecule in the orthorhombic lattice in Ref.~\onlinecite{Curran24}.

\section{Theoretical Model}

We consider the case where  triplet excitons can localize at either one of two inequivalent sites, which can be visualized as the two families of molecules with different orientations in Fig.~\ref{Axes}. In the main-axes cartesian coordinate system that diagonalizes the spin Hamiltonian for an individual exciton at one site, the stationary state energies at zero magnetic field for an exciton at that site can be written in terms of the zero-field splitting parameters $D$ and $E$ \cite{Yarmus72,Swenberg73,Clarke76}. The spin-dependent Hamiltonian for an individual triplet exciton is 
\begin{equation}
   {H_T} = D\left( {S}_z^2-\frac{2}{3}\right)+E\left( {S}_x^2- {S}_y^2\right), \label{ZFS}
\end{equation}
where ${S}_i$ are the components of the exciton spin vector, with $S^2=2$. In the main-axes coordinate system of a given site, as an example the $x_A, y_A, z_A$ coordinate system in Fig.~\ref{Axes}, the spin states of a triplet exciton at that site can be represented by the spin wavefunctions $\ket{x_A}, \ket{y_A}, \ket{z_A}$, which are also eigenstates of the Hamiltonian at zero magnetic field (the spin operator and Hamiltonian commute) and are labeled by the coordinate axes for which the corresponding spin component vanishes. 

In order to study a triplet pair with excitons that can localize at different sites, we use the global coordinate system $x,y,z$ shown in Fig.~\ref{Axes}, and the  basis of  product states $\ket{xx}, \ket{xy}, \ket{xz},\ket{yx}, \ket{yy}, \ket{yz}, \ket{zx}, \ket{zy}, \ket{zz}$. In the limit of negligible triplet-triplet interactions, and at zero-field, each product state has a corresponding energy that is the sum of individual triplet energies ($E_{i,j}=E_i+E_j$). In the following, we will neglect triplet-triplet interactions between two excitons in a pair on the grounds that  the dipole-dipole interaction  decays with the cube of the distance, that it  is orders of magnitude smaller than the difference between stationary state energies \cite{Wang15}, and that we focus on the evolution of the triplet-pair state in a lattice where the two excitons can separate by diffusion.

The zero-field Hamiltonians for a triplet pair depend on its spatial configuration, namely, which site each triplet exciton in the pair occupies. We use $A$ and $B$ labels for the two inequivalent sites and categorize triplet exciton pairs based on the type of site where each exciton is found. In this way we have ``same-type'' pairs in $AA$ or $BB$ spatial configurations, and ``mixed-type'' pairs in a $AB$ spatial configuration. The configuration-dependent Hamiltonians are 
\begin{eqnarray}
  H_{con}^{(AA)} &= &H_A \otimes I + I \otimes H_A, \\
  H_{con}^{(BB)} &= &H_B \otimes I + I \otimes H_B, \\
  H_{con}^{(AB)} &=& H_A \otimes I + I \otimes H_B,
\end{eqnarray}
where $H_A$ and $H_B$ are the individual triplet Hamiltonians for each site, the symbol $\otimes$ represents the Kronkecker Product, and $I$ is the identity operator. $H_A$ and $H_B$ are in general obtained by a unitary coordinate transformation from the main-axes system of each site onto the desired global coordinate system, such as  $x,y,z$, in Fig.~\ref{Axes} for rubrene crystals. In this case, $H_A$ and $H_B$ are obtained by a simple $\pm \theta$ rotation  of the Hamiltonian defined in the main-axes system of an inequivalent site.

The magnetic-field Hamiltonian for a state with spin $\vec S$ is $H_{Z} = g \mu_B \vec S \cdot \vec B$, where $\vec B$ is the magnetic field, $g \approx 2$ is the gyromagnetic factor, and $\mu_B$ is the Bohr magneton. 

The total Hamiltonian for the triplet pair is $H = H_{con} + H_{mag}$, with $H_{mag} = H_Z \otimes I + I \otimes H_Z$, and it depends both on the triplet-pair spatial configuration and on the magnetic field strength and direction.

When a triplet pair is generated from a photoexcited singlet state, and angular momentum conservation applies, the pair is entangled into an overall singlet state, given in the zero-field basis of the product states by
\begin{equation}
  \ket{S} = \frac{1}{\sqrt{3}}\bigl[\ket{x_i x_i}+\ket{y_i y_i}+\ket{z_i z_i}\bigr], \label{AASinglet}
\end{equation}
where $i$ specifies one of the three coordinate systems in Fig.~\ref{Axes}. Since the spin operator and Hamiltonian do not commute in general, this overall singlet state is not a stationary state. And the spin-wavefunctions on the right-hand side of (\ref{AASinglet}) are in general not stationary states as well. They can be stationary states only at zero magnetic field, when both excitons occupy the same type of site, and when the coordinates are expressed in the main-axes coordinate system of that site. 

\begin{table*}
\caption{\label{StationaryStatesAB} Zero-field orthonormal stationary states $\psi_i$ for  a triplet pair in the $AB$ configuration for the rubrene example of Fig.~\ref{Axes}.  The stationary states are given both in the  mixed-coordinate basis described in the text, and in the basis that uses the global coordinate system $x,y,z$ in Fig.~\ref{Axes}. The corresponding eigenenergies and singlet projection amplitudes $\braket{S}{\psi_i}$ are  also provided. The last two lines of the table give expressions for the overall triplet-pair singlet state: First, as a function of the four stationary states that contribute to it, and then as a function of the basis states in the mixed and global coordinates.}
\begin{ruledtabular}
\begin{tabular}{lcllc}
& Energy & Mixed-Coordinate Basis & Global-Coordinate Basis & $\braket{S}{\psi_i}$ \\
\colrule
$\psi_1$& $2(D/3+ E)$ & $\ket{z_A z_B}$ & $ \phantom{-} \ket{zz}$ & $ \frac{1}{\sqrt{3}}$ \\
$\psi_2$& $2(D/3- E)$ & $\ket{y_A y_B}$ & $ -\ket{xx} \sin ^2\theta +\ket{yy} \cos ^2\theta + \frac{1}{2}(\ket{xy}-\ket{yx})\sin(2 \theta )$ & $ \frac{1}{\sqrt{3}} \cos (2 \theta)$ \\
$\psi_3$& $-\frac{4 D}{3}$ & $\ket{x_A x_B}$ & $ +\ket{xx} \cos ^2\theta -\ket{yy} \sin ^2\theta + \tfrac{1}{2}(\ket{xy}-\ket{yx})\sin(2 \theta)$ & $ \frac{1}{\sqrt{3}} \cos (2 \theta)$ \\
$\psi_4$& $-D/3-E$ & $\frac{1}{\sqrt{2}}(\ket{x_A y_B}-\ket{y_Ax_B})$ & $ - \frac{1}{\sqrt{2}} (\ket{xx}+\ket{yy})\sin (2 \theta) + \frac{1}{\sqrt{2}}(\ket{xy}-\ket{yx})\cos (2 \theta) $ & $ \sqrt{\frac{2}{3}} \sin (2 \theta)$ \\
$\psi_5$& $-D/3-E$ & $\frac{1}{\sqrt{2}}(\ket{x_A y_B}+\ket{y_Ax_B})$ & $ \phantom{-}\tfrac{1}{\sqrt{2}} [\ket{xy}+\ket{yx}]$ & $ 0 $ \\
$\psi_6$& $-D/3+E$ & $\frac{1}{\sqrt{2}}(\ket{x_A z_B}-\ket{z_Ax_B})$ & $ -\frac{1}{\sqrt{2}}(\ket{xz}-\ket{zx})\cos(\theta) + \frac{1}{\sqrt{2}}(\ket{yz}+\ket{zy})\sin(\theta)$ & $ 0 $ \\
$\psi_7$& $-D/3+E$ & $\frac{1}{\sqrt{2}}(\ket{x_Az_B}+\ket{z_Ax_B})$ & $\phantom{-}\frac{1}{\sqrt{2}}(\ket{xz}+\ket{zx})\cos(\theta) - \frac{1}{\sqrt{2}}(\ket{yz}-\ket{zy})\sin(\theta)$ & $ 0 $ \\
$\psi_8$& $2 D/3$ & $\frac{1}{\sqrt{2}}(\ket{y_A z_B}-\ket{z_Ay_B})$ & $ - \frac{1}{\sqrt{2}}(\ket{yz}-\ket{zy})\cos(\theta) - \frac{1}{\sqrt{2}}(\ket{xz}+\ket{zx})\sin(\theta)$ & $ 0$ \\
$\psi_9$& $2 D/3$ & $\frac{1}{\sqrt{2}}(\ket{y_A z_B}+\ket{z_Ay_B})$ & $\phantom{-} \frac{1}{\sqrt{2}}(\ket{yz}+\ket{zy})\cos(\theta) + \frac{1}{\sqrt{2}}(\ket{xz}-\ket{zx})\sin(\theta) $ & $ 0 $ \\
 \\
$\ket{S}_{AB}$ & 0 & \multicolumn{2}{l}{ \hspace{0.6cm}
$
\tfrac{1}{\sqrt{3}} \left( \psi_1 + \cos(2 \theta)(\psi_2+\psi_3) - \sqrt{2} \sin(2 \theta)\psi_4 \right)$ 
} & 1 \\
$\ket{S}_{AB}$& 0 & Eq.~(\ref{ABSinglet})& $\tfrac{1}{\sqrt{3}}[\ket{xx}+\ket{yy}+\ket{zz}]$ & $ 1$\\
\end{tabular}
\end{ruledtabular}
\end{table*}

For a triplet pair created in the overall singlet state, the probability that triplet-triplet annihilation at a later time $t$ can result in an emissive singlet state is the singlet projection probability
\begin{equation}
P_S(t) = \left|\langle S|U(t)|S\rangle \right|^2 ,
\label{UexpVal}
\end{equation}
where $U(t)$ is the unitary time-evolution operator for the system. 
For a time-independent Hamiltonian $H = H_{con} + H_{mag}$, the time-evolution operator is
\begin{equation} 
U(t) = \exp \left( - i H t/ \hbar \right). \label{Uoperator}
\end{equation}

A time-independent Hamiltonian would apply when no inequivalent sites are present or when the spatial wavefunction of the triplet exciton is larger than the distance between inequivalent sites. In such a case Eq.~(\ref{UexpVal}) leads to well-defined quantum interference upon pair recombination (triplet exciton fusion), observable as fluorescence quantum beats after pulse photoexcitation and singlet fission \cite{Funfschilling85b,Chabr81,Burdett12,Wolf18}. For a constant Hamiltonian and in the absence of a magnetic field,  the overall singlet state is in general a superposition of three stationary states with different energies, leading to quantum beats that contain three frequencies, as observed in tetracene crystals \cite{Funfschilling85b,Chabr81,Burdett12}.

However, in the presence of exciton transport, the spatial configuration of the exciton pair and the corresponding Hamiltonian change as the excitons hop between inequivalent sites. While the configuration Hamiltonian for each site is time-independent, the triplet-pair Hamiltonian becomes a function of time. 
The time-evolution operator must then be written as
\begin{equation}
U(t) = \prod_{n=1}^{N} U_{n}(\Delta t_n) \label{inotaryRandomWalk}
\end{equation}
where $N$ is the total number of configuration changes that are expected during the time $t$, $\Delta t_n$ are the dwell-times in each configuration, $U_n(t) = \exp \left( - i H_{n} t/ \hbar \right)$, and $H_{n} = H_{con,n} + H_{mag}$. Here, $H_{con,n} = H_{con}^{(AB)}$ when $n$ is even, and $H_{con,n} = H_{con}^{(AA)}$ or $H_{con,n} = H_{con}^{(BB)}$, with equal probability, when $n$ is odd ($n=1$ corresponds to the configuration created by singlet fission).

In the limit of fast hopping, when hopping rates are very large compared with any energy differences between triplet-pair stationary states, the product in Eq.~(\ref{inotaryRandomWalk}) reverts to Eq.~(\ref{Uoperator}), but  \change{with a time-independent Hamiltonian $H = H_{eff} + H_{mag}$ where the zero-field part is  $H_{eff} =  (H_{con}^{(AA)} + H_{con}^{(BB)} + 2 H_{con}^{(AB)})/4$, the average over the triplet-pair Hamiltonians in the different pair configurations, weighted by the relative dwell times. In this limit, the effective Hamiltonian has in general four stationary states with singlet projections (similar to the $H_{AB}$ Hamiltonian  in Table \ref{StationaryStatesAB}), which then  leads  to zero-field quantum beats that contain more than 3 frequencies. The convergence to   $H_{eff} =  (H_{con}^{(AA)} + H_{con}^{(BB)} + 2 H_{con}^{(AB)})/4$ in the fast hopping limit is confirmed by the full Monte Carlo computation presented below. } 

It is important to stress that for triplet excitons that can localize at inequivalent sites it would be incorrect to evaluate the fast hopping limit by deriving  a single product-state triplet-pair Hamiltonian from the site-averaged Hamiltonian of an individual exciton  (as might be obtained from EPR measurements \cite{Yarmus72} or ab-initio calculations). On the other hand,  a triplet-wavefunction  larger than the distance between inequivalent sites would justify such an approach. This feature  may potentially be helpful to experimentally distinguish between the cases of a large exciton wavefunction and fast hopping. As an example, using the $H_{eff}$ given above to calculate tetracene quantum beats in the fast hopping limit and at zero field does indeed lead to more than three beat frequencies, with the lowest frequency near 1 GHz splitting into a doublet separated by about 0.1$-$0.2 GHz. The fact that   only one frequency at about 1.06 GHz is observed experimentally \cite{Burdett12} may imply that for tetracene one has  a large exciton wavefunction instead of just fast hopping, but this interpretation is constrained by the limited precision of the available tetracene zero-field parameters \cite{Yarmus72,Clarke76} and experimental data \cite{Burdett12}.

The general case where the hopping times are larger than a few picoseconds  cannot be described by a single effective Hamiltonian anymore, and Eq.~(\ref{inotaryRandomWalk}) must be used. We compute this product of time-evolution operators using a Monte Carlo simulation of the random walk between spatial triplet-pair configurations, with a sequence of dwell times $\Delta t_n$ that follow an exponential distribution $\tau^{-1} \exp(-t/\tau)$. Here, $\tau = \tau_{hop}/2$ is the average triplet-pair dwell time in any specific configuration, equal to half the average hopping time $\tau_{hop}$ of an individual triplet exciton between inequivalent sites.

\section{Quantum Beat Behavior}

\begin{figure*}
  \includegraphics[width=0.9\textwidth]{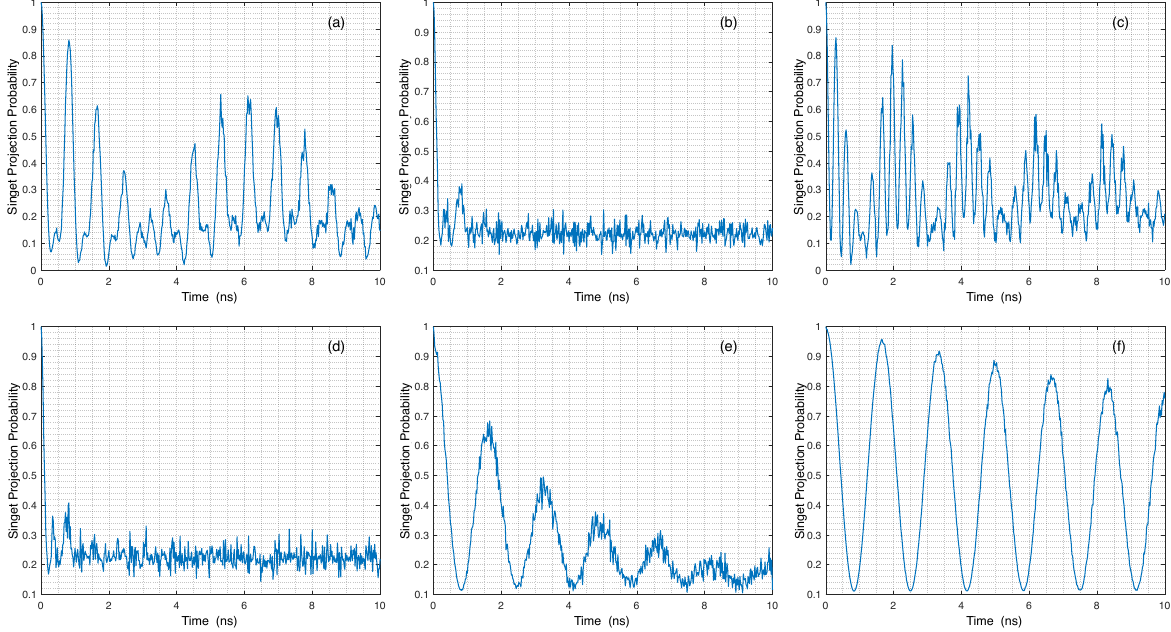}
  \caption{Time-dependent Singlet Projection Probabilities $P_S(t)$ (Eq.~\ref{UexpVal}) for the rubrene crystal geometry and for different exciton hopping times and the zero-field parameters of Ref.~\onlinecite{Curran24}. The top row shows quantum beats predicted with no magnetic field for $x$-axis exciton hopping times $\tau_{hop}$ of 5 ps ($a$), 150 ps ($b$), and 10 ns ($c$). The bottom row displays predicted quantum beats for $\tau_{hop} = 150$~ps and an external magnetic  field  of magnitudes 0 T (d), 0.3 T (e), and 1.0 T (f) and parallel to  the global $y$-axis. Transport-induced dephasing effects \cite{Curran24} do not occur with this orientation of magnetic field.} \label{Sims}
\end{figure*}

Fig.~\ref{Sims} shows the time-dependence of the Singlet-Projection Probabilities of Eq.~(\ref{UexpVal}), predicted using Eqs.~(\ref{inotaryRandomWalk}) and (\ref{UexpVal}) in a Monte Carlo computation with exponentially distributed dwell times, rubrene molecular parameters \cite{Curran24}, and a choice of exciton hopping times along the $x$-axis that vary over 4 orders of magnitude from $5$ ps to $10$ ns. The $x$-axis in the global coordinate system (Fig.~\ref{Axes}) is the direction of exciton diffusion that causes a change from a same-type ($AA$ or $BB$) to a mixed-type ($AB$) triplet-pair; $y$-axis diffusion moves the exciton along the molecular column with no change in spatial configuration. The singlet-projection probabilities of Fig.~\ref{Sims} directly give the quantum beats that can be extracted \cite{Wolf18} from the short-pulse induced fluorescence in rubrene crystals.

Fig.~\ref{Sims}($a$-$c$) shows zero-field quantum beats. 
For a ``fast'' hopping time of the order of picoseconds, one has an effective average configuration Hamiltonian $H_{eff}= [H_{con}^{(AA)}+ H_{con}^{(BB)} + 2 H_{con}^{(AB)}]/4$, as mentioned above, leading to zero-field quantum beats similar to what is observed in tetracene fluorescence  \cite{Burdett12}. \change{In this case one observes the persistent quantum beats in Fig.~\ref{Sims}($a$), which can be shown to be a superposition of oscillations with 5 different frequencies of 0.16, 1.15, 1.31, 2.29, and 2.45 GHz.}

On the other hand, for a ``slow'' hopping time of the order of 10 ns (Fig.~\ref{Sims}$c$), $x$-axis hopping is so slow that triplet-pair spatial configurations are unlikely to change, and one observes quantum beats that are initially determined by the  configuration Hamiltonian $H_{con}^{(AA)}$ or $H_{con}^{(BB)}$. Since the probability of moving to an $AB$ configuration remains low on the time scale of Fig.~\ref{Sims}, this results in quantum beats that are initially a superposition of three oscillations with  the three frequencies of a same-type pair: 0.48, 3.09, and 3.57 GHz.  

At intermediate hopping times (Fig.~\ref{Sims}$b$), the variability of the dwell time in any given configuration (equal to the average value itself for exponentially distributed times) becomes important. A significant number of configuration changes with a large variance around the mean \cite{Curran24} leads to an effective randomization of the evolution of the triplet-pair spin wavefunction, in a way that is described by the product of time-evolution operators in Eq.~(\ref{inotaryRandomWalk}). This results in a rapid destruction of global coherence in the triplet-pair population. At zero magnetic field, and in the example of rubrene, the most severe zero-field decoherence occurs for a hopping time between inequivalent sites  in the range of about 100 to 200 ps, in agreement with the high-field results of Ref.~\onlinecite{Curran24}. In this range  our treatment predicts  either no quantum beats at zero field, or quantum beats that decay completely within 1 ns, to a level corresponding to a singlet projection probability of 2/9.

It is interesting to note that this analysis, based solely on zero-field experiments, delivers an estimate for the average hopping time that is consistent with that found in Ref. \onlinecite{Curran24} by studying a high-field dephasing effect. 

Next, Fig.~\ref{Sims}($d$-$f$) shows the quantum-beat behavior once an external magnetic field is applied in a direction symmetric with respect to the inequivalent sites, the geometry that eliminates the transport-induced dephasing effect that was discussed in Ref.~\onlinecite{Curran24}. Application of a magnetic field in this way makes the corresponding Zeeman Hamiltonian increasingly dominant, leading to essentially single-frequency quantum beats (frequency of 0.61 GHz) and allowing global coherence to recover. But the part of the triplet-pair Hamiltonian that depends on the triplet-pair configuration still leads to global decoherence because of random hopping between inequivalent sites. At low magnetic field strengths this leads to quantum beats with a quickly decaying amplitude, with the steady-state singlet projection probability reaching $1/9$ when the magnetic field has a component perpendicular to $z$ (as in Fig.~\ref{Sims}($d$-$f$)), and  $2/9$ when the magnetic field is along the $z$ direction. 

The final average value of the singlet-projection probability after full decoherence---of 2/9 at zero-field and for a magnetic field in the $z$-direction, or 1/9 otherwise---is related to the fact that in the rubrene example of Fig.~\ref{Axes} the $\ket{zz}$ component of the pair wavefunction remains invariant both at zero-field and for magnetic field along $z$, but it suffers from hopping-induced decoherence otherwise. When global decoherence is complete, the probability of any triplet-pair to still be in an overall singlet state decays to $1/9$ (corresponding to a random distribution of possible spin states), but the absence of a magnetic field along $x$ or $y$, and the particular crystal structure of rubrene where all molecules have a main axis parallel to $z$,  helps maintain partial coherence.

As the external field becomes stronger, zero-field decoherence is increasingly suppressed, and the quantum-beat amplitude stabilizes, reaching predicted decay times larger than $60$ ns for a magnetic field of $1$ T in the $z$-direction.

\section{Discussion}

The results presented above show that our Monte Carlo treatment of the triplet-pair configuration changes due to hopping can predict transport-induced global decoherence for any hopping time between inequivalent sites, and any applied magnetic field. The results are consistent with earlier experimental investigations of quantum beats in rubrene \cite{Wolf18,Curran24}.

\change{At zero-field, the recovery of quantum beats in the limit of fast hopping, which is displayed in Fig.~\ref{Sims}($a$), is reminiscent of  motional line narrowing, in a manner similar to what was discussed in the high field limit in Ref.~\onlinecite{Curran24}, when analyzing the   global dephasing that causes a decay of the quantum-beat amplitude when the magnetic field direction does not exactly align with a symmetry direction. This is naturally related to the fact that quantum-beat frequencies can be more precisely determined for beats that persist for a longer time. In general, it is better to focus on the decay time of the beat amplitude (or of the average singlet projection probability) as the key quantity that can be directly determined experimentally.}

More insights into the origin of transport-induced global decoherence can be gained by considering how the stationary states of the system vary with the spatial configuration of the entangled triplet pair. To repeat the key point here, stochastic triplet-exciton hopping causes the  triplet pair to repeatedly transition to and from the $AB$ configuration. It is therefore interesting to discuss the corresponding mixed-type triplet-pair in more detail. 

Table~\ref{StationaryStatesAB} gives all the  zero-field stationary states for the mixed-type  $AB$ configuration of the triplet-pair. For any stationary state $\psi$, we can obtain useful insights by considering the singlet projection probabilities $|\braket{S}{\psi}|^2$ and the corresponding singlet projection amplitudes $\braket{S}{\psi}$, which are also given in Table~\ref{StationaryStatesAB}. These singlet projections affect the probability of triplet-triplet fusion and its time-dependence, both in the case of entangled triplet pairs, and more generally for an unentangled triplet population \cite{Merrifield71,Swenberg73}. 

As shown in Table~\ref{StationaryStatesAB}, the mixed-type $AB$ triplet pair  has four stationary states with singlet projections at zero field, instead of just the three states $\ket{xx}$, $\ket{yy}$, and $\ket{zz}$ found in  same-type triplet pairs. 
The fourth state, $\psi_4$ in Table~\ref{StationaryStatesAB}, is degenerate with the symmetric state $\psi_5$, but the latter does not contribute to the overall singlet wavefunction.

Table~\ref{StationaryStatesAB} also lists the stationary states using a ``mixed-coordinate'' basis for the spin-eigenstates of the two excitons, where the spin of each exciton is defined in its own main-axes system, as $\ket{i_A}$ or $\ket{i_B}$, and the spin of the exciton pair is described by $\ket{i_A j_B}$ ($i,j=x,y,z$).
For zero magnetic field, and neglecting triplet-triplet interactions, these mixed-coordinate states for the $AB$ configuration are automatically stationary states of the system. The result of applying the Hamiltonian to them is simply 
\begin{equation}
H_{con} \ket{i_A}\ket{j_B} = (E_i + E_j) \ket{i_A}\ket{j_B} \label{ABEnergy},
\end{equation}
where $i,j = x,y,z$, and $E_i$ and $E_j$ are the eigenenergies of the spin states of each exciton in the pair in its own main-axes system.

In the example of Fig.~\ref{Axes}, one can relate these states to those defined in the main-axes system of site $A$, by a simple rotation of only the elements with $B$ subscript around the $z$-axis by the angle $2 \theta$:
\begin{align*}
  \ket{x_A x_A} &= + \cos(2\theta)\ket{x_A x_B} - \sin(2\theta)\ket{x_A y_B} \\
  \ket{x_A y_A} &= + \sin(2\theta)\ket{x_A x_B} + \cos(2\theta)\ket{x_A y_B} \\
  \ket{x_A z_A} &= \ket{x_A z_B} \\
  \ket{y_A x_A} &= + \cos(2\theta)\ket{y_A x_B} - \sin(2\theta)\ket{y_A y_B} \\
  \ket{y_A y_A} &= + \sin(2\theta)\ket{y_A x_B} + \cos(2\theta)\ket{y_A y_B} \\
  \ket{y_A z_A} &= \ket{y_A z_B} \\
  \ket{z_A x_A} &= + \cos(2\theta)\ket{z_A x_B} - \sin(2\theta)\ket{z_A y_B} \\
  \ket{z_A y_A} &= + \sin(2\theta)\ket{z_A x_B} + \cos(2\theta)\ket{z_A y_B} \\
  \ket{z_A z_A} &= \ket{z_A z_B}
\end{align*}

The  triplet-pair singlet state, which in the main-axes system of one site is $(\ket{x_A x_A}+\ket{y_A y_A}+\ket{z_A z_A})/\sqrt{3}$, then becomes
\begin{multline}
  \ket{S}_{AB} = \frac{1}{\sqrt{3}} \big[ \cos(2\theta)(\ket{x_A x_B}+\ket{y_A y_B}) + \ket{z_A z_A}\big] \\ 
  - \sqrt{\frac{2}{3}}\sin(2\theta)\frac{1}{\sqrt{2}}(\ket{x_A y_B} -\ket{y_A x_B}) \label{ABSinglet}
\end{multline}
in the mixed-coordinate basis. When describing a triplet pair in the $AB$ configuration, the  mixed-coordinate basis consists of stationary spin states but the overall singlet wavefunction must be written as the linear combination of the four basis states in Eq.~\ref{ABSinglet}. On the other hand, in the global ($x,y,z$) basis, the  zero-field stationary states are  linear combinations of several spin states, as seen in  Table~\ref{StationaryStatesAB}, but the overall singlet wavefunction  is a simple sum of three spin states (Eq.~\ref{AASinglet}) that applies to  all triplet-pair configurations

In any case, both Eq.~(\ref{ABSinglet}) and Table~\ref{StationaryStatesAB} deliver identical singlet projections, which indeed must be independent from basis states and the choice of coordinate systems. 

This gives a new view into the unique characteristics of the $AB$ configuration.
The last term in Eq.~(\ref{ABSinglet}) shows that the normalized antisymmetric state $(\ket{x_A y_B} - \ket{y_A x_B})/\sqrt{2}$ contributes to the singlet wavefunction with a significant singlet projection probability that increases with the rotation angle between inequivalent sites and becomes strongest when the two local main-axes systems are perpendicular to each other. In rubrene ($\theta = 31^\circ$), this fourth state has a singlet projection probability of 0.52. As noted earlier, this state is degenerate with the corresponding symmetric state, which has zero singlet projection.

\begin{figure}
  \includegraphics[width=0.9\columnwidth]{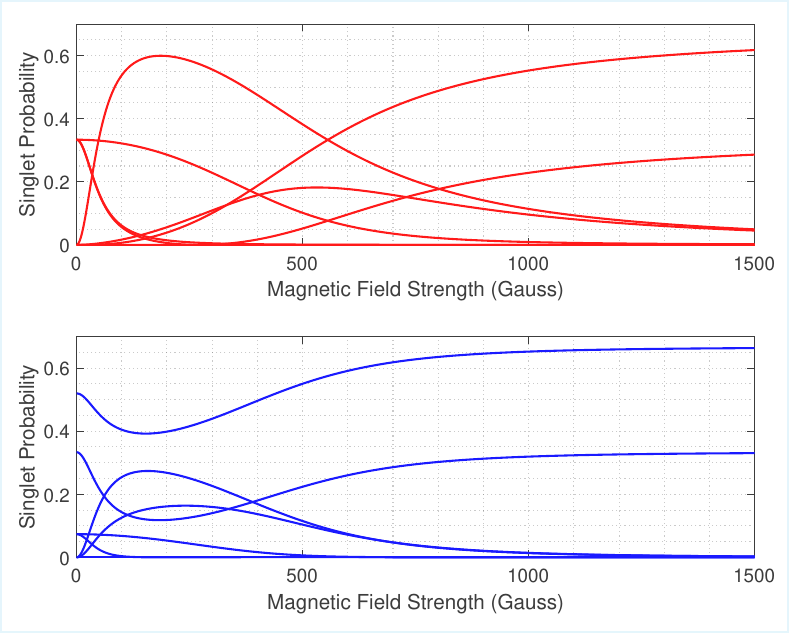} \\
  \caption{Singlet projection probabilities for each of the nine stationary states for same-type triplet pairs ($AA$/$BB$ configuration, top, red) and mixed-type pairs ($AB$ configuration, bottom, blue), as a function of the strength of a magnetic field applied along the $x$-axis.}
  \label{MATLAB}
\end{figure}

A qualitative picture of the zero-field behavior of the triplet pair in the presence of transport and its effect on global decoherence and  observable quantum beats can be painted as follows. At zero-field, the spin wavefunction of a same-type triplet pair is the same for both $AA$ and $BB$ configuration, with three stationary states that contribute to the overall singlet state and are in general non-degenerate, leading to a time-dependent singlet projection characterized by three frequencies. On the other hand, a mixed-type triplet pair  in the $AB$ configuration has four non-degenerate stationary states with singlet contributions, corresponding to six frequencies. The zero-field decoherence effect can then be seen to arise from continued transitions between each triplet-pair configuration, with random dwell times, which ultimately leads to mixing of all the spin states but for $\ket{zz}$.

\begin{table}[h!]
\caption{\label{tabStationaryStates} Energies (in GHz) and singlet projection probabilities of the stationary states contributing to the overall singlet state of a triplet pair in the rubrene structure.}
\begin{tabular}{ccr}
\hline\hline
State & $E/h$ (GHz) & $|\langle S|\psi\rangle|^2$ \\
\hline
\multicolumn{3}{c}{\it{Zero Field, AA configuration}} \\
$|z_A z_A\rangle \equiv |zz\rangle$ & 0.87 & $0.3\overline{3}$ \\
$|y_A y_A\rangle$ & 1.35 & $0.3\overline{3}$ \\
$|x_A x_A\rangle$ & -2.22 & $0.3\overline{3}$\vspace{6pt} \\

\hline
\multicolumn{3}{c}{\it{Zero Field, AB configuration}} \\
$|z_A z_B\rangle \equiv |zz\rangle$ & 0.87 & $0.3\overline{3}$ \\
$|y_A y_B\rangle$ & 1.34 & 0.07 \\
$|x_A x_B\rangle$ & -2.21 & 0.07 \\
$\frac{1}{\sqrt{2}}(|x_A y_B\rangle - |y_A x_B\rangle)$ & -0.43 & 0.52\vspace{6pt} \\
\hline
\multicolumn{3}{c}{\it{High Field, magnetic field along $x$-axis}} \\
$\frac{1}{\sqrt{2}}(\ket{1,-1}+\ket{-1,1})$ & 0.64 & $0.6\overline{6}$ \\
$|0,0\rangle$ & -1.27 & $0.3\overline{3}$\vspace{6pt} \\
\hline\hline
\end{tabular}
\end{table}

When a magnetic field is applied, the stationary states change, and so do their singlet projection probabilities. This is shown for the example of rubrene in Fig.~\ref{MATLAB}, where singlet projection probabilities for each stationary state are plotted as a function of magnetic field strength for both the same-type and the mixed-type triplet pairs. In general, the number of stationary states that contribute to the singlet state increases when a moderate magnetic field below 0.1 T is applied, which is then followed by the expected convergence to just two states with a non-vanishing singlet projection in the high-field limit. Numerical values for zero-field and high-field singlet projection and stationary state energies for the example of rubrene are listed in Table~\ref{tabStationaryStates}.

The difference between the same-type triplet pair and the mixed-type one is again notable.
As seen in Table~\ref{tabStationaryStates}, the $AA$/$BB$ configuration starts with three equal-probability singlet projections at zero field, while the $AB$ configuration starts with four states with singlet projections at zero field, with the largest singlet projection belonging to the state $\psi_4 =(\ket{x_A y_B} - \ket{y_A x_B})/\sqrt{2}$. This state then evolves into a high-field state that can be written as $(\ket{1,-1} + \ket{-1,1})/\sqrt{2}$ in terms of the spin-components along the magnetic field, and which contributes to the singlet state at high field, which is given by $((\ket{1,-1} + \ket{-1,1}) - \ket{0,0})/\sqrt{3}$ regardless of triplet-pair configuration. In rubrene, this leads to the expected single-frequency high-field quantum beats \cite{Wolf18, Curran24} whenever transport-induced dephasing is avoided by ensuring that the Zeeman Hamiltonian is the same for both inequivalent sites \cite{Curran24}. 

But as shown in Fig.~\ref{Sims}, even in this case where transport-induced \emph{dephasing} does not occur, transport-induced \emph{decoherence} still leads to a decay of quantum-beat amplitude. In contrast to high-field transport-induced dephasing \cite{Curran24}, this decoherence effect does not depend on quantum-beat frequency. In fact, while quantum beats for magnetic fields along the $x$ and $y$ direction have different frequencies (1.88 and 0.61 GHz, respectively, in rubrene \cite{Curran24}), the corresponding amplitude decay times are the same for both magnetic field orientations (about 4 ns exponential decay time, as shown for $y$-oriented magnetic field in Fig.~\ref{Sims}($e$)). On the other hand, transport-induced decoherence effects are weaker for magnetic fields oriented along the $z$-direction, with quantum-beat decay times that are much longer (by more than a factor of 3) and higher final singlet projection probability that is caused by the invariance of the $\ket{zz}$ component of the spin-wavefunction in the example of the rubrene structure. This is related to the fact that for this orientation the magnetic field is exactly along the same main axis for both inequivalent sites in Fig.~\ref{Axes}, which makes the $\ket{zz}$ state a stationary state for all magnetic field strengths and for both same-type and mixed-type triplet pairs.

In this context, it should be noted that another unique property of the $AB$ triplet-pair configuration is the fact that the states $\ket{-1,1}$ and $\ket{1,-1}$ can still differ by a fraction of the zero-field energy up to relatively high magnetic fields whenever the magnetic field has components along both $x$ and $y$. This was shown in Ref.~\onlinecite{Tapping16}, where an $AB$-configuration of the triplet pair was calculated to have 3 stationary states, each with singlet probability of $1/3$, up to relatively large fields. But in such a situation one should build symmetric and antisymmetric linear combinations of the two quasi-degenerate high-field stationary states, which then again leads to essentially two high-field states with singlet probabilities of $2/3$ and $1/3$.

\section{Conclusion}
We presented a complete analysis of global decoherence in a triplet-pair population where triplet excitons can localize at different inequivalent sites in a crystal lattice, and developed a model that can predict fluorescence quantum beats and their properties for any value of magnetic field and exciton hopping time between inequivalent sites. We have shown that a simple Monte-Carlo approach can be used to take into account the stochastic changes in triplet-pair configuration caused by random hopping of the excitons in a pair, and to obtain an effective unitary time-evolution operator.

We also discussed the properties of triplet-pairs in different spatial configurations, and highlighted the fact that one more stationary state strongly contributes to the overall singlet state at zero field when the excitons in a pair reside on different sites.

The model can be applied in general to any crystal structure with two inequivalent sites. Results derived for the example of rubrene single crystals are consistent with earlier quantum-beat experiments \cite{Wolf18,Curran24}. In particular, we have shown that our analysis predicts the absence of  quantum beats at zero magnetic field in rubrene for a range of hopping times between inequivalent sites that is consistent with an earlier estimate obtained from an analysis of a different transport induced dephasing effect in the high-field limit \cite{Curran24}.

\begin{acknowledgments}
Research supported by the US Department of Energy, Office of Basic Energy Sciences, Division of Materials Sciences and Engineering, under Award No.~DE-SC0020981.
\end{acknowledgments}

\bibliography{RubreneBib}

\end{document}